\documentclass[10pt,conference]{IEEEtran}
\usepackage{epsfig}
\usepackage{amsmath}
\usepackage{amsfonts}
\usepackage{amssymb}
\usepackage{graphicx}
   \DeclareGraphicsExtensions{.eps}

\newcommand{\gf}{{\mathbb F}}%
\newcommand{\mbf}[1]{{\mathbf{#1}}}%
\newcommand{\dx}{\mbox{\rm d}x}%
\newlength{\tmplength}%
\newcommand{\tspace}[1][4.5mm]{\settoheight{\tmplength}{X}\addtolength{\tmplength}{#1}\hbox{\protect\raisebox{\tmplength}{}}}
\newenvironment{mylist}[1][$\bullet$]{\begin{list}{#1}{\leftmargin \parindent \itemindent 0mm \labelwidth \parindent}}{\end{list}}

\IEEEoverridecommandlockouts
\title{Optimized puncturing distributions for irregular non-binary LDPC codes\vspace{-2mm}}%
\author{Matteo~~Gorgoglione$^{\ast}$, Valentin~Savin$^{\ast}$, David~Declercq$^{\natural}$\\
   $^{\ast}$CEA-LETI, MINATEC, Grenoble, France, \{valentin.savin, matteo.gorgoglione\}@cea.fr \\
   $^{\natural}$ETIS ENSEA/univ. Cergy-Pontoise/CNRS, Cergy-Pontoise, France, declercq@ensea.fr%
\vspace{-2mm}%
\thanks{This work was carried out in the scope of the APOGEE
project,
supported by the French Research and Innovation Program for Telecommunication.
}}%
\date{}
\begin{document}
\maketitle

\begin{abstract}
\, In this paper we design non-uniform bit-wise puncturing distributions for irregular non-binary LDPC (NB-LDPC) codes. The puncturing distributions
are optimized by minimizing the decoding threshold of the punctured LDPC code, the threshold being computed with a Monte-Carlo implementation
of Density Evolution. First, we show that Density Evolution computed with Monte-Carlo simulations provides accurate (very
close) and precise (small variance) estimates of NB-LDPC code ensemble thresholds. Based on the proposed method, we
analyze several puncturing distributions for regular and semi-regular codes, obtained either by clustering punctured
bits, or spreading them over the symbol-nodes of the Tanner graph. Finally, optimized puncturing distributions for non-binary
LDPC codes with small maximum degree are presented, which exhibit a gap between $0.2$ and $0.5$ dB to the channel
capacity, for punctured rates varying from $0.5$ to $0.9$.
\end{abstract}


\section{Introduction}
In the modern approach of error correcting codes, binary Low-Density Parity-Check (LDPC) codes are playing a very important role
due to their low decoding complexity and to the fact that they can be optimized for a broad class of channels, with
asymptotic performance approaching the theoretical Shannon limit \cite{Rich-Shok-Urba}. On the other hand, LDPC codes
over non-binary alphabets exhibit better performance for short to medium code lengths \cite{Davey-MacKey}, without sacrifying the
almost optimal asymptotic performance. The performance gain at short length comes at an additionnal cost in the decoding complexity.
However, this additionnal cost can be reduced by using sub-optimal low complexity algorithms, as proposed e.g. in \cite{Decl-Foss}, \cite{savin_min_max}.
When coded bits are transmitted over time-varying channels, as for wireless communication systems, the rate of the forward error correction code has to be
adapted to the quality of the channel, so as to achieve both high throughput and reliability. Rate adaptation requires
codes of different rates, which can be efficiently obtained by using one low rate mother code and puncture it to
get higher rates. The advantage of puncturing is that the same decoder can be used regardless the puncturing pattern:
according to the channel conditions, the transmission system adapts by just changing the
puncturing pattern at the transmitter, and the depuncturing pattern at the receiver.

In this work we explore the rate-adaptability behavior of non-binary LDPC codes. In the binary case, Ha et al.
\cite{Ha-Kim-McLaug} showed that puncturing patterns can be optimized so that rate-compatible puncturing results in a
small loss of the asymptotical performance (threshold). For non-binary codes, Klinc et al. \cite{Klinc-Ha-McLaug}
proposed a design of puncturing patterns based on symbol recovery levels, similar to the design proposed in the
binary case in \cite{Ha-Kim-Klinc-McLaug}. Given a target punctured rate, the algorithm proposed in \cite{Ha-Kim-Klinc-McLaug} computes a puncturing
pattern that minimizes the number of recovery steps for the punctured bits. This approach is thought for finite length
puncturing design, but does not give insights regarding the optimization of the puncturing distribution, for LDPC code ensembles.

In the binary case, the puncturing distribution is
defined by fractions $\{f_d\}_d$ of punctured bit-nodes of degree $d$. For non-binary codes, the puncturing
distribution can be defined in terms of fractions $\{f_{d,k}\}_{d,k}$ of degree-$d$ symbol-nodes with exactly $k$
punctured bits per symbol. Note that addressing the problem of the optimization of puncturing distributions is compliant
with the finite length approach of \cite{Klinc-Ha-McLaug}, and from an optimized puncturing distribution, one can
advantageously use the algorithm proposed in \cite{Klinc-Ha-McLaug} in order to construct the actual puncturing pattern
\footnote{As a parallel, consider the construction of bipartite Tanner graphs: for given node-degree
distributions, we can advantageously use the Progressive Edge Growth (PEG) algorithm, so as to maximize the girth of
the constructed graph; however, if the PEG algorithm is not constrained with a given degree distribution, the algorithm will
tend to construct a regular graph, which does not provide the best performance}.

Although the study of non-binary LDPC codes is mainly motivated by their performance at short to medium lengths, their
asymptotic analysis proves also to be very useful. In particular, it allows the understanding of topology independent
phenomena or, stated differently, it allows distinguishing between events due to code family parameters, and those due to a
particular code realization.
The asymptotical performance of an ensemble of codes is referred to as {\em threshold}, and corresponds to the worst
channel condition that allows transmission with an arbitrary small error probability when the code length tends to
infinity. It can be determined by the concept of density evolution \cite{Rich-Urba}. Unfortunately, for the non-binary case,
exact density evolution is only manageable for the Erasure Channel \cite{rathi-Urbanke}, \cite{savin2008nbl}. In the case of
more general memryless channels, like the
binary-input AWGN channel, density evolution must be approximated using numerical techniques, or approched by modeling the density
of the messages, e.g. with a Gaussian approximation \cite{Li-Fair-Krzy}.

In this work we show that density evolution of non-binary LDPC codes can be tightly approximated by using fast
Monte-Carlo approximation of the density of the messages exchanged during the iterative decoding of an infinite code.
This method allows for obtaining very close estimates of thresholds of non-binary ensembles, making possible the optimization of
non-binary codes for a wide range of applications and channel models. In particular, it can be successfully applied for
optimizing puncturing distributions for rate-compatible non-binary LDPC codes. We used genetic optimization methods for
searching good puncturing distributions for non-binary LDPC codes over $\gf_{16}$, with maximum symbol-node degree
equal to $10$. Optimized  distributions exhibit a gap to capacity between $0.2$ and $0.5$dB, for punctured rates
varying from $0.5$ to $0.9$.

The rest of the paper is organized as follows. In Section \ref{sec:nbldpc} we introduce the notations and review some
basics on non binary LDPC codes. Section \ref{sec:mcde_nbldpc} presents the proposed method for approximating the
density evolution of non-binary LDPC codes. Puncturing distributions are introduced in Section \ref{sec:punct_distrib},
and clustering vs. spreading distributions are analyzed for regular and semi-regular codes in Section
\ref{sec:punct_analysis}. The optimization of puncturing distributions for irregular codes is addressed in Section
\ref{sec:punct_optimization}. Finally, Section \ref{sec:conclusions} concludes the paper.

\section{Non binary LDPC codes} \label{sec:nbldpc}
We consider non-binary codes defined over $\gf_q$, the finite field with $q$ elements, where $q = 2^p$ is a power of
$2$ (the last condition is only assumed for practical reasons). We fix once for all a vector space isomorphism:
\begin{equation}
    \label{identify}
     {\gf_q} \stackrel{\sim}{\longrightarrow} \gf_2^p,
\end{equation}
and we say that $(x_0,\dots,x_{p-1})\in\gf_2^p$ is the {\em binary image} of the {\em symbol} $X\in{\gf_q}$ if they
correspond to each other by the above isomorphism. An LDPC code over $\gf_q$ is defined as the kernel of a sparse
parity-check matrix $H \in \mbf{M}_{M,N}(\gf_q)$. The matrix $H$ can be represented by a bipartite graph comprising $M$
{\em constraint-nodes} and $N$ {\em symbol-nodes}, associated respectively with its $M$ rows and its $N$ columns. A
constraint-node and a symbol-node are connected by an edge if the corresponding entry of $H$ is non-zero. Each edge of
the graph is further labeled by the corresponding non-zero entry. Thus, symbol-nodes represent coded symbols, while
constraint-nodes together with edge labels correspond to linear constraints on these symbols. The number of edges
incident to a node is referred to as node degree.
Interconnections between nodes are usually expressed by the edge-perspective degree distribution polynomials
$\lambda(x) = \sum_{d}\lambda_d x^{d-1}$ and $\rho(x) = \sum_{d}\rho_d x^{d-1}$, where $\lambda_d$ and $\rho_d$
represent represent the fraction of edges connected respectively to symbol and constraint-nodes of degree $d$. We
denote by $r$ the design coding rate, which is given by $r = 1 - \frac{\int_0^1 \rho(x) \,\dx}{\int_0^1 \lambda(x)
\,\dx}$. It is well known that the asymptotic performance of binary LDPC codes depends only on these degree
distribution polynomials [8]. For non-binary codes, the performance can be slightly improved by choosing optimized
distributions of edge labels [10]. However, in general this improvement seems to be minimal; therefore, in the next
section we will consider ensembles ${\cal E}_q(\lambda, \rho)$, of non-binary LDPC codes over $\gf_q$, with
edge-perspective degree distribution polynomials $\lambda$ and $\rho$, and with edge labels uniformly distributed over
$\gf_q^{*}$.

\section{Asymptotic performance of non-binary LDPC codes}\label{sec:mcde_nbldpc}
\begin{figure}[!b]
 \vspace{-5mm}\centering
 \includegraphics*[height=.95\linewidth,angle=-90]{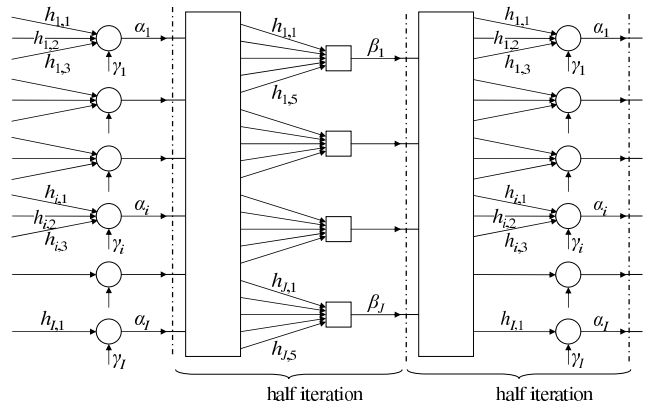}
 \caption{Density evolution approximation by Monte-Carlo simulation}\label{fig:MC_DE}
 \vspace{-3mm}
\end{figure}

Consider some channel model depending on a parameter $s$, such that the channel conditions worsen when $s$ increases
(for instance, the noise variance for the AWGN channel, or the erasure probability for the BEC channel). The {\em
threshold}\break of the ensemble ${\cal E}_q(\lambda, \rho)$ is defined as the supremum value of $s$ (worst channel
condition) that allows transmission with an arbitrary small error probability, assuming that the transmitted data is
encoded with an arbitrary-length code of ${\cal E}_q(\lambda, \rho)$. For symmetric channels, we can assume that the
all-zero codeword is transmitted, and the threshold can be approximated by simulating the density evolution of messages
exchanged within the iterative decoding of an ``infinite'' code from the ensemble \cite{Davey-McKay:inf_nbldpc}. For
doing so, we have to ensure that exchanged messages remain uncorrelated for a large number of iterations. Since the
density evolution is observed on a finite number of exchanged messages, we proceed as described below in order to
decorrelate them. Consider that we have to compute an outgoing message (either from a symbol, or a constraint-node).
First, we sample the node degree $d$ according to the corresponding degree distribution (either $\lambda$, or $\rho$),
and we randomly choose $d-1$ messages from those produced at the previous half iteration. These messages are considered
as being received by the node on the $d-1$ incoming edges, and $d-1$ edge labels are also uniformly sampled from
$\gf_q^{*}$. Besides, in order to compute outgoing messages from symbol-nodes, we also need the knowledge the a priori
information corresponding to the channel output. In this case, for each new outgoing symbol-node message, the channel
output is also resampled according to the channel model. The outgoing message is then computed according to the
Sum-Product decoding rules \cite{Wiberg}. This is illustrated at Figure \ref{fig:MC_DE} for $\lambda(x) = \frac{1}{3}x
+ \frac{2}{3}x^3$ and $\rho(x) = x^5$ (designed code rate $r=1/2$), where outgoing symbol and constraint-node messages
are respectively denoted by $\alpha_i$ and $\beta_j$, edge labels are denoted by $h_{i,j}$, and the a priori
information corresponding to the channel output by $\gamma_i$. We also  remark  that  exchanged  messages  are better
decorrelated for higher order finite fields, due to the edge-labels resampling.

Table \ref{tab:thresh_bin_compare} presents a comparison of thresholds obtained by the proposed method (MC-DE) and
those obtained by exact density evolution, for irregular binary\footnote{Exact DE for non-binary LDPC codes over the
AWGN channel is not known.} LDPC codes with rate $1/2$ over the AWGN channel. Thresholds are are shown both in terms of
noise variance and corresponding $E_b/N_0$ value.  The maximum bit-node degree is reported in the left column; the
exact degree distributions $\lambda$ and $\rho$ can be found in \cite{Rich-Shok-Urba}, Tables I and II. For comparison
purposes, we have also included threshold values computed by using the Gaussian-Approximation proposed in
\cite{Chung-Richar-Urba:DE_GA}. At each half-iteration of the Monte-Carlo Density-Evolution (MC-DE) simulation, $10000$
outgoing messages are computed, and a maximum number of $500$ iterations are executed. For each pair of degree
distributions $(\lambda, \rho)$, several threshold estimations have been performed; the mean value is reported in the
MC-DE/$\sigma^{*}$ column, and the standard deviation ($\times 10^3$) is reported in the last column. As it can be
observed, the standard deviation is between $0.0008$ and $0.0017$; hence we conclude that the proposed method is
precise. Furthermore, the Root Mean Square Error (RMSE) of estimated thresholds (reported on the last row) if equal to
$0.0031$, which proves that the proposed method is also accurate.

\begin{table}[!t]
\caption{Thresholds for binary irregular LDPC codes}\label{tab:thresh_bin_compare} \footnotesize \noindent
  \begin{tabular}{|c||c|@{\,}c@{\,}||c|@{\,}c@{\,}||c|@{\,}c@{\,}|c@{\,}|}
  \hline
    $d_{\mbox{\scriptsize max}}$   &  \multicolumn{2}{c||}{Exact-DE} & \multicolumn{2}{|c||}{GA-DE} & \multicolumn{2}{|c|}{MC-DE} & $\sigma^{*}_{\mbox{\scriptsize dev}}$\\
    \cline{2-7}
    &  $\sigma^{*}$ & $E_b/N_0^{*}$ &  $\sigma^{*}$ & $E_b/N_0^{*}$ &  $\sigma^{*}$ & $E_b/N_0^{*}$ &  $\times 10^3$  \\
    \hline
  4  &   0.911 & 0.808  &   0.904 & 0.876  &  0.916 & 0.757 & 1.2 \\
  5  &   0.919 & 0.730  &   0.911 & 0.806  &  0.923 & 0.691 & 1.1 \\
  6  &   0.930 & 0.627  &   0.907 & 0.774  &  0.931 & 0.622 & 1.0 \\
  7  &   0.942 & 0.515  &   0.922 & 0.699  &  0.935 & 0.577 & 1.7 \\
  8  &   0.950 & 0.448  &   0.938 & 0.557  &  0.947 & 0.468 & 1.5 \\
  9  &   0.954 & 0.409  &   0.940 & 0.533  &  0.954 & 0.411 & 1.1 \\
 10  &   0.956 & 0.393  &   0.942 & 0.523  &  0.955 & 0.403 & 1.2 \\
 11  &   0.957 & 0.380  &   0.942 & 0.518  &  0.957 & 0.377 & 1.1 \\
 12  &   0.958 & 0.373  &   0.942 & 0.513  &  0.958 & 0.367 & 0.8 \\
 15  &   0.962 & 0.335  &   0.944 & 0.503  &  0.962 & 0.340 & 1.6 \\
 20  &   0.965 & 0.310  &   0.946 & 0.482  &  0.963 & 0.325 & 0.7 \\
 30  &   0.969 & 0.273  &   0.946 & 0.469  &  0.965 & 0.299 & 1.6 \\
 50  &   0.972 & 0.248  &   0.952 & 0.428  &  0.971 & 0.260 & 0.8 \\
 \hline
 \multicolumn{3}{|c||}{RMSE} & 0.0168 & 0.1459 & 0.0031 & 0.0272 & \multicolumn{1}{c}{}\\
 \cline{1-7}
  \end{tabular}
  \vspace{-3mm}
\end{table}

\section{Puncturing distributions for non-binary LDPC codes}\label{sec:punct_distrib}
\begin{figure}[!t]
\noindent
 \includegraphics*[height=\linewidth,angle=-90]{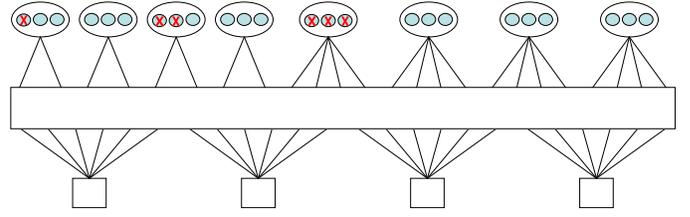}
 \caption{Bit level puncturing pattern for non-binary LDPC codes}\label{fig:punct}
 \vspace{-4mm}
\end{figure}

Let a non binary LDPC code be used over a binary-input channel: a non-binary codeword is mapped into its binary images
by using (\ref{identify}), then transmitted over the channel. Hence, even if non-binary codes operate at the symbol
level, non-binary codewords still can be punctured at the bit-level. A coded symbol can be either completely or
partially punctured, according to whether all the bits of its binary image are punctured or not. In order to design
good puncturing patterns we have to answer the following questions:
\begin{mylist}
\item Assuming that the code is regular, should puncturing bits be clustered or spread over the symbol-nodes?
Clustering yields a reduced number of completely punctured symbols, which receive no information from the channel. On
the contrary, spreading results into an increased number of partially punctured symbols, which still benefit from some
partial information from the channel.
\item In case of irregular codes, how should punctured bits be chosen with respect to symbol-node degrees?
\end{mylist}
Let $f_{d,k}$ denote the fraction of degree-$d$ symbol-nodes with $k$ punctured bits.
Note that $\sum_{k=0}^p f_{d,k} = 1$, as any symbol-node contains either $k=0$ (unpunctured), or $k=1$, $\dots$, or
$k=p$ punctured bits. The overall fraction of punctured bits, $f$, and the corresponding punctured rate, $r_p$, are
given by:
        $$f = \frac{1}{p}\sum_{d=1}^{d_s}\sum_{k=0}^p k f_{d,k} L_d, \ \ r_p = \frac{r}{1-f}$$
where $d_s$ is the maximum symbol-node degree, and $L_d = \displaystyle\frac{\lambda_d}{d\int_0^1 \lambda(x)\,\dx}$ is
the fraction of symbol-nodes of degree $d$. This is illustrated at Figure \ref{fig:punct} for an LDPC code over $\gf_8$
with designed code rate $r=1/2$, where $f_{2,1} = f_{2,2} = 1/4$, and $f_{4,3} = 1/4$. Thus $f =0.25$ and the designed
punctured rate is $r_p = 2/3$.

\section{Analysis of puncturing distributions}\label{sec:punct_analysis}
In this section we analyze different puncturing distributions for regular and semi-regular\footnote{With symbol-node
degrees taking on two, relatively low, values.} non-binary LDPC codes over the BIAWGN channel.

\begin{figure}
\includegraphics*[width=\linewidth]{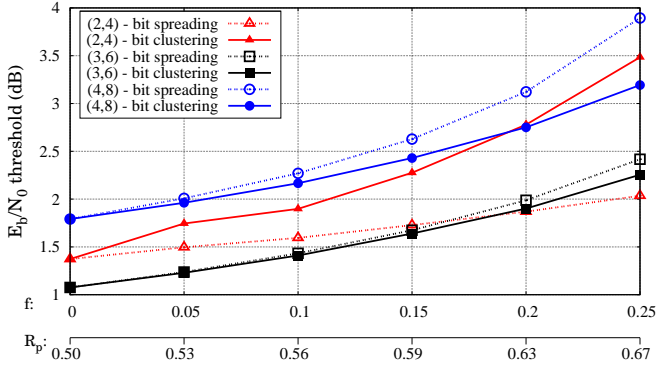}
\vspace{-8mm} \caption{Spreading vs. clustering distribution; regular LDPC codes over $\gf_8$}\label{fig:gf8_r12}
\vspace{-3mm}
\end{figure}

\begin{figure}
\includegraphics*[width=\linewidth]{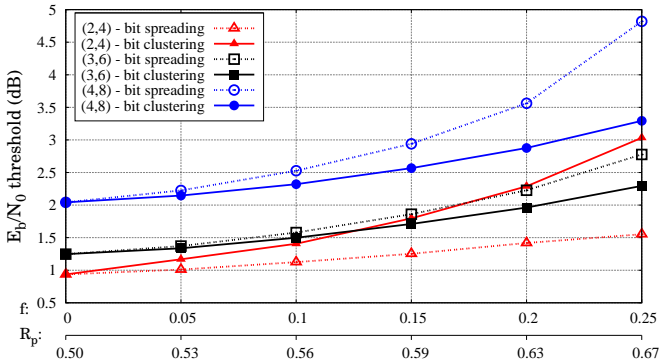}
\vspace{-8mm}\caption{Spreading vs. clustering distribution; regular LDPC codes over $\gf_{16}$}\label{fig:gf16_r12}
\vspace{-3mm}
\end{figure}

First of all, we consider three ensembles of regular LDPC codes with rate $1/2$, and (symbol, constraint)-node degrees
equal to $(2, 4)$, $(3, 6)$, or $(4, 8)$. A fraction $f\in[0, 0.25]$ of coded bits are punctured, which corresponds to
a punctured rate varying from $r_p = 1/2$ (no puncturing) to $r_p = 2/3$. For each fraction of punctured bits, two
puncturing distributions have been compared:
\begin{mylist}
\item clustering distribution: punctured bits are clustered on a fraction $f$ of completely punctured symbols-nodes;
\item spreading distribution: if $f<\frac{1}{p}$ punctured bits are spread over a fraction $pf$ of symbol-nodes, each one with one single
punctured bit. Otherwise, all the symbol-nodes are punctured, each symbol-node containing either $\lfloor pf \rfloor$
or $\lceil pf \rceil$ punctured bits
\end{mylist}
Corresponding thresholds for codes over $\gf_8$ and $\gf_{16}$ are shown respectively at Figure \ref{fig:gf8_r12} and
Figure \ref{fig:gf16_r12}. We note the different behavior of these distributions, depending on the symbol-node degree:
the spreading distribution provides better performance for the regular $(2, 4)$ code, but it is outperformed by the
clustering distribution for the other two codes with higher symbol-node degrees.

The intuition behind that goes as follows. An outgoing message from a degree-$2$ symbol-node depends on the information
coming from the channel and the unique incoming extrinsic message. Thus, if a degree-$2$ symbol-node is completely
punctured, it has no channel information, and the outgoing message is equal to the incoming one. This causes a lack of
diversity in the iterative message-passing, which results in degraded decoding performance.

\begin{figure}
\includegraphics*[width=\linewidth]{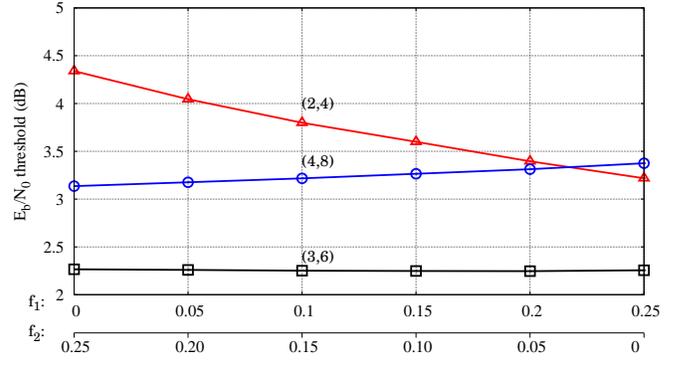}
\vspace{-8mm}\caption{Intermediary puncturing distributions for regular codes over $\gf_4$}\label{fig:gf4_r12}
\vspace{-3mm}
\end{figure}

\begin{figure}[!t]
\includegraphics*[width=\linewidth]{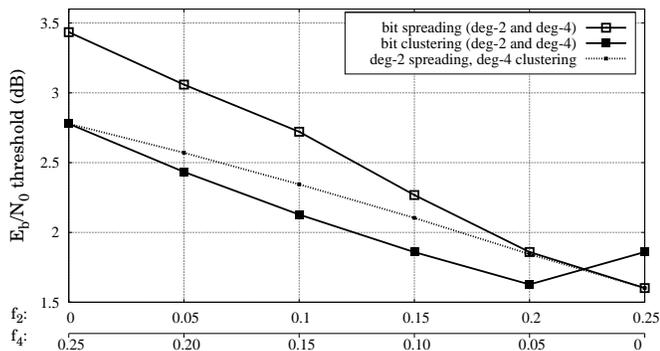}
\vspace{-8mm}\caption{Low vs. high symbol-degree,  and spreading vs. clustering puncturing distributions for irregular
LDPC codes over $\gf_16$}\label{fig:GF16_SEMI_REG_r067}\vspace{-3mm}
\end{figure}

Intermediary puncturing distributions can be obtained by mixing the above clustering and spreading distributions, as
illustrated at Figure \ref{fig:gf4_r12} for regular codes over $\gf_4$: the fraction of punctured bits is $f=0.25$,
corresponding to a punctured rate $r_p = 2/3$, and it is decomposed as $f=f_1+f_2$, where $f_1$ is the fraction of bits
that are punctured in a spreading manner (in this case, $1$ bit per symbol), and $f_2$ is the fraction of bits that are
punctured in a clustering manner ($2$ bits per symbol). Again, we can observe that the spreading distribution provides
better performance for the regular $(2,4)$ code. Similar results are also shown at figure \ref{fig:GF16_SEMI_REG_r067}
for codes over $\gf_{16}$: for a clusterization degree $k\in\{1,2,3,4\}$ (on the abscissa), a fraction $\frac{1}{k}$ of
symbol-nodes are punctured, by punching $k$ bits per symbol. Once more, we can observe that spreading is suitable for
symbol-nodes of degree-$2$, while clustering is more appropriate for symbol-nodes of degree $\geq 3$.

We investigate now different puncturing distributions for semi-regular LDPC codes, with node-perspective degree
distribution polynomials $L(x) = \frac{1}{2}x^2 + \frac{1}{2}x^4$ and $R(x) = x^6$, corresponding to a designed coding
rate $r=1/2$. The same fraction $f=0.25$ of coded bits are punctured, so as to obtain a punctured rate $r_p=2/3$. The
fraction is decomposed as $f=f_2+f_4$, where $f_2$ is the fraction of bits that are punctured on degree-$2$
symbol-nodes, and $f_4$ is the fraction of bits that are punctured on degree-$4$ symbol-nodes.

\begin{figure}
\includegraphics*[width=\linewidth]{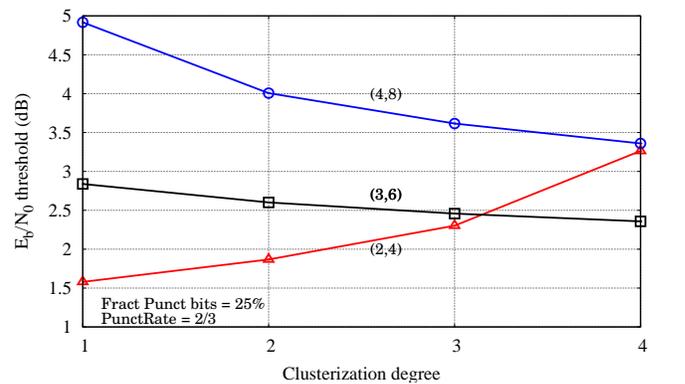}
\vspace{-8mm}\caption{Various clusterization degrees for regular codes over $\gf_{16}$ for
}\label{fig:gf16_r12_f25_r067} \vspace{-3mm}
\end{figure}

\begin{figure}[!t]
\includegraphics*[width=\linewidth]{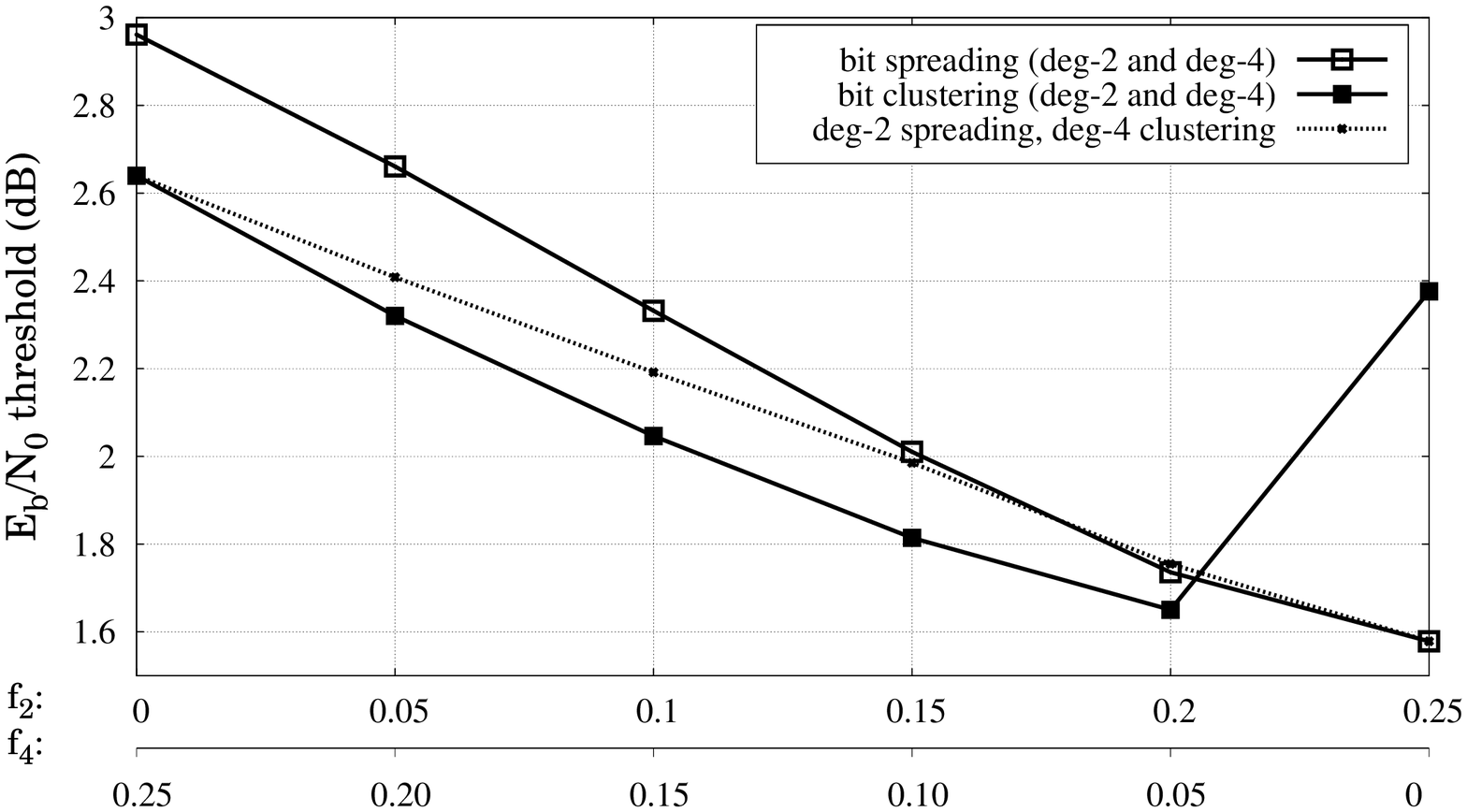}
\vspace{-8mm}\caption{Low vs. high symbol-degree, and spreading vs. clustering puncturing distributions for irregular
LDPC codes over $\gf_8$}\label{fig:GF8_SEMI_REG_r067} \vspace{-3mm}
\end{figure}

Numerical results are shown at Figure \ref{fig:GF8_SEMI_REG_r067} and Figure \ref{fig:GF16_SEMI_REG_r067}: for solid
curves, punctured bits are either clustered (full squares) or spread over the symbol-nodes (empty squares), for both
symbol-nodes of degree $2$ and $4$. For the dashed curve, punctured bits are spread over symbol-nodes of degree $2$,
and clustered over symbol-nodes of degree $4$. In view of previous results for regular codes, we was expecting this
last distribution to provide the best performance. Surprisingly, it performs in between the clustering and the
spreading distribution. What we observe is that in case there are punctured symbols of degree $4$ ($f_4 > 0$), the best
performance is given by the clustering distribution. However, since all plots correspond to the same punctured rate,
the best puncturing distribution in each figure is given by the lowest plot. In both cases this is the distribution
that spreads all the punctured bits over symbol-nodes of degree $2$ ($f_2 = 0.25$, $f_4=0$, empty
square)\footnote{However, the clustering distribution with $f_2=0.2$ and $f_4 = 0.05$ yields almost the same
performance}. This represents an important dissimilarity compared to the binary case, in which choosing all the
punctured bits on degree-$2$ bit-nodes proves to be catastrophic. In the non-binary case, spreading punctured bits over
degree-$2$ symbol nodes yields good performance, {\em provided that the fraction of punctured bits is not too high}.
The next section will also provide evidence on this fact.

\section{Optimization of puncturing distributions}\label{sec:punct_optimization}

\begin{figure}[!t]
\includegraphics*[width=\linewidth]{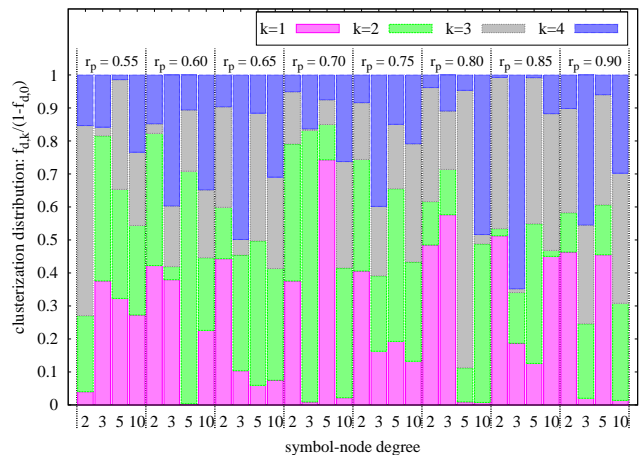}
\vspace{-6mm}\caption{Clusterization distributions for $r_p\in\{0.55,\dots,0.90\}$}\label{fig:clusterization_distrib}
\vspace{-3mm}
\end{figure}

In this section we present optimized puncturing distributions for an irregular code over $\gf_{16}$ with rate $1/2$,
and punctured code rates from $0.55$ to $0.9$. Optimization has been performed by using the Differential Evolution
algorithm \cite{Storn-Price}. First of all we searched for good degree distribution pairs $\lambda$ and $\rho$ of rate
$1/2$, with maximum symbol degree\footnote{Cf. discussion from the Introduction.} equal to $10$. The optimized
distributions and the corresponding $E_b/N_{0}$ threshold are given below:
$$\begin{array}{rcl}
\lambda(x) & = & 0.5376 x + 0.1678 x^2 + 0.1360 x^4 + 0.1586 x^9 \\
\rho(x) & = & 0.5169 x^4 +   0.4831 x^5 \\
E_b/N_{0}^{*} & = & 0.3638
\end{array}$$
Next, we searched for good puncturing distributions for punctured rates $r_p\in\{0.55, 0.60, 0.65, 0.70, 0.75, 0.80,
0.85, 0.90\}$. Optimized distributions $\{f_{d,k}\}_{\stackrel{\mbox{\tiny $d=2,3,5,10$}}{\mbox{\tiny $k=0,1,2,3,4$}}}$
for punctured rates $0.60$, $0.75$, and $0.90$ are shown in Table \ref{tab:opt_distrib}. Also indicated are
$\bar{k}_d$, the average number of punctured bits per symbol-node of degree $d$, and $f_k$, the average fraction of
symbols with $k$ punctured bits. As an equivalent representation, the corresponding {\em clusterization distributions}
are shown at Figure~\ref{fig:clusterization_distrib}: they consist of fractions of {\em punctured
symbols}\footnote{Recall that $f_{d,k}$ is the fraction of {\em symbols} of degree-$d$, with $k$ punctured bits per
symbol} of degree-$d$, with $k$ punctured bits per symbol, which are given by $\frac{f_{d,k}}{1-f_{d,0}}$, for
$d=2,3,5,10$, and $k=1,\dots,4$. These distributions seem to be random, and properties observed for regular codes shall
not apply is this case.

\begin{table}[!t]
\caption{Optimized distribution, $r_p = 0.60, 0.75, 0.90$}\label{tab:opt_distrib}
 \begin{center}\footnotesize
 \vspace{-3mm}
\begin{tabular}{|@{\,}c@{\,}|ccccc|c|}
\hline
\tspace[1.3mm]$r_p= 0.60$ &  $k = 0$ &  $k=1$   &  $k=2$   &  $k=3$   &  $k=4$   & $\bar{k}_d$\\
\hline
 $d=2$ &  0.7132  &  0.1211  &  0.1149  &  0.0083  &  0.0425  &  0.5456 \\
 $d=3$ &  0.5953  &  0.1536  &  0.0161  &  0.0743  &  0.1607  &  1.0514 \\
 $d=5$ &  0.7414  &  0.0009  &  0.1822  &  0.0479  &  0.0275  &  0.6190 \\
$d=10$ &  0.4608  &  0.1218  &  0.1187  &  0.1109  &  0.1878  &  1.4429 \\
\hline
 $f_k$ &  0.6865  &  0.1172  &  0.1050  &  0.0257  &  0.0656  &         \\
\hline \hline
\tspace[1.3mm]$r_p= 0.75$ &  $k = 0$ &  $k=1$   &  $k=2$   &  $k=3$   &  $k=4$    & $\bar{k}_d$\\
\hline
 $d=2$  &  0.2026  &  0.3865  &  0.1047  &  0.2757  &  0.0308  &  1.5448 \\
 $d=3$  &  0.3616  &  0.3680  &  0.0876  &  0.1129  &  0.0699  &  1.1615 \\
 $d=5$  &  0.5783  &  0.0038  &  0.0435  &  0.3546  &  0.0197  &  1.2335 \\
$d=10$  &  0.1998  &  0.0060  &  0.3841  &  0.0230  &  0.3870  &  2.3913 \\
\hline
 $f_k$  &  0.2545  &  0.3390  &  0.1096  & 0.24585  &  0.0510  &         \\
\hline \hline
\tspace[1.3mm]$r_p= 0.90$ &  $k = 0$ &  $k=1$   &  $k=2$   &  $k=3$   &  $k=4$   & $\bar{k}_d$\\
\hline
 $d=2$  & 0.0960  &  0.4187  &  0.1077  &  0.2857  &  0.0919  &  1.8589 \\
 $d=3$  & 0.6543  &  0.0070  &  0.0779  &  0.1035  &  0.1572  &  1.1023 \\
 $d=5$  & 0.1304  &  0.3957  &  0.1314  &  0.2905  &  0.0521  &  1.7384 \\
$d=10$  & 0.0413  &  0.0132  &  0.2822  &  0.3780  &  0.2854  &  2.8530 \\
\hline
$f_k$ & 0.1811  &  0.3369  &  0.1125  &  0.2623  &  0.1072  &         \\
\hline
\end{tabular}
\end{center}
\vspace{-3mm}
\end{table}

\begin{figure}[!t]
\includegraphics*[width=\linewidth]{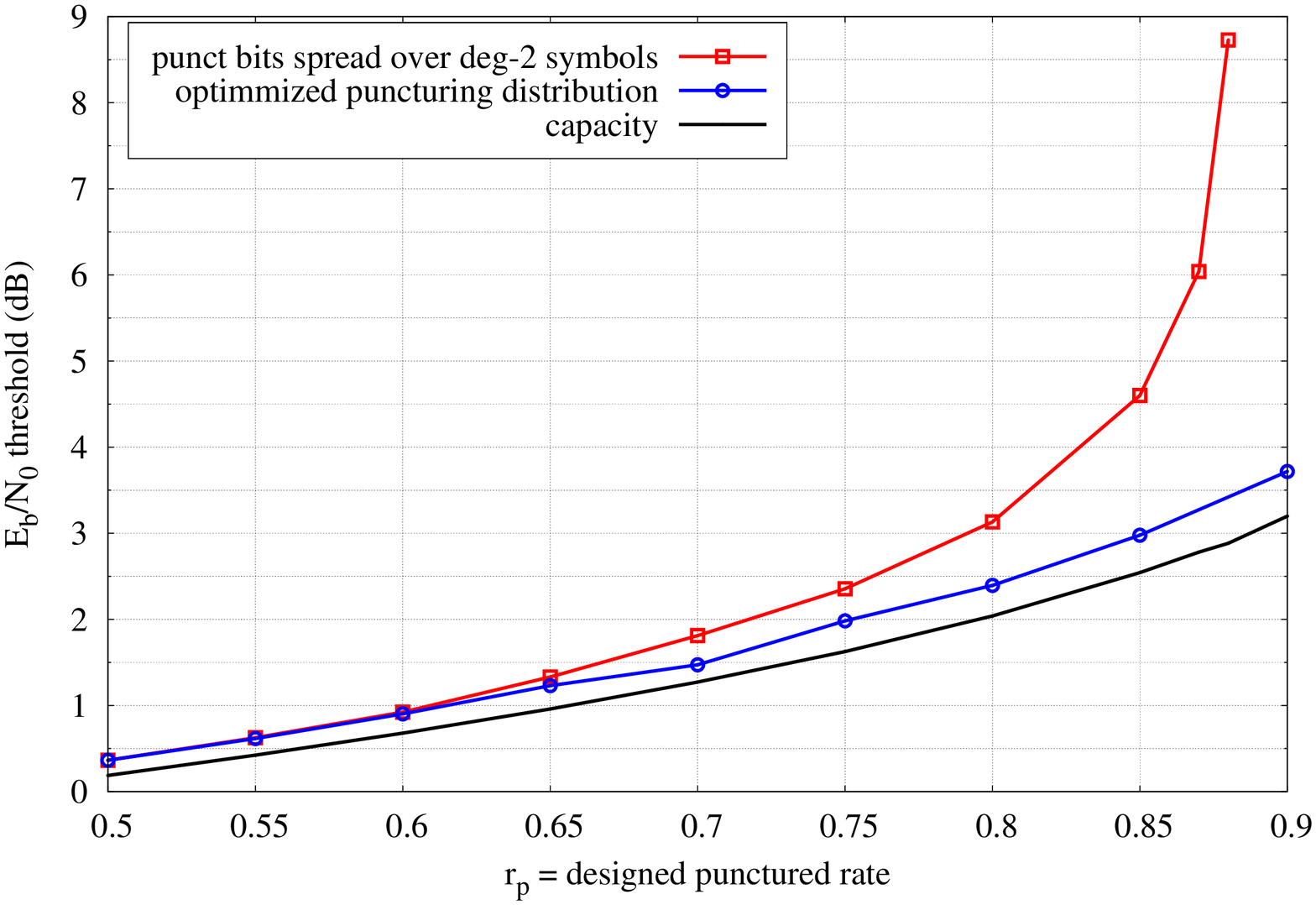}
\vspace{-8mm}\caption{Optimized puncturing distributions for regular codes over $\gf_{16}$}\label{fig:gf16_opt_distrib}
\vspace{-3mm}
\end{figure}

Thresholds of optimized punctured distributions are shown at Figure \ref{fig:gf16_opt_distrib}, in terms of $E_b/N_0$.
Are also plotted the theo\-retical Shannon limit (capacity), and thresholds of distributions spreading punctured bits
over degree-$2$ symbol-nodes. The gap between optimized distributions and capacity vary between $0.18$ db for $r_p =
0.5$ (unpunctured code) and $0.52$ dB for $r_p = 0.9$. We also note that the degree-$2$-spreading distribution yields
very good performance up to punctured rate $r_p = 0.7$. However, such a puncturing distribution proves to be
catastrophic for punctured rates $r_p \geq 0.85$

\begin{figure}[!t]
\includegraphics*[width=\linewidth]{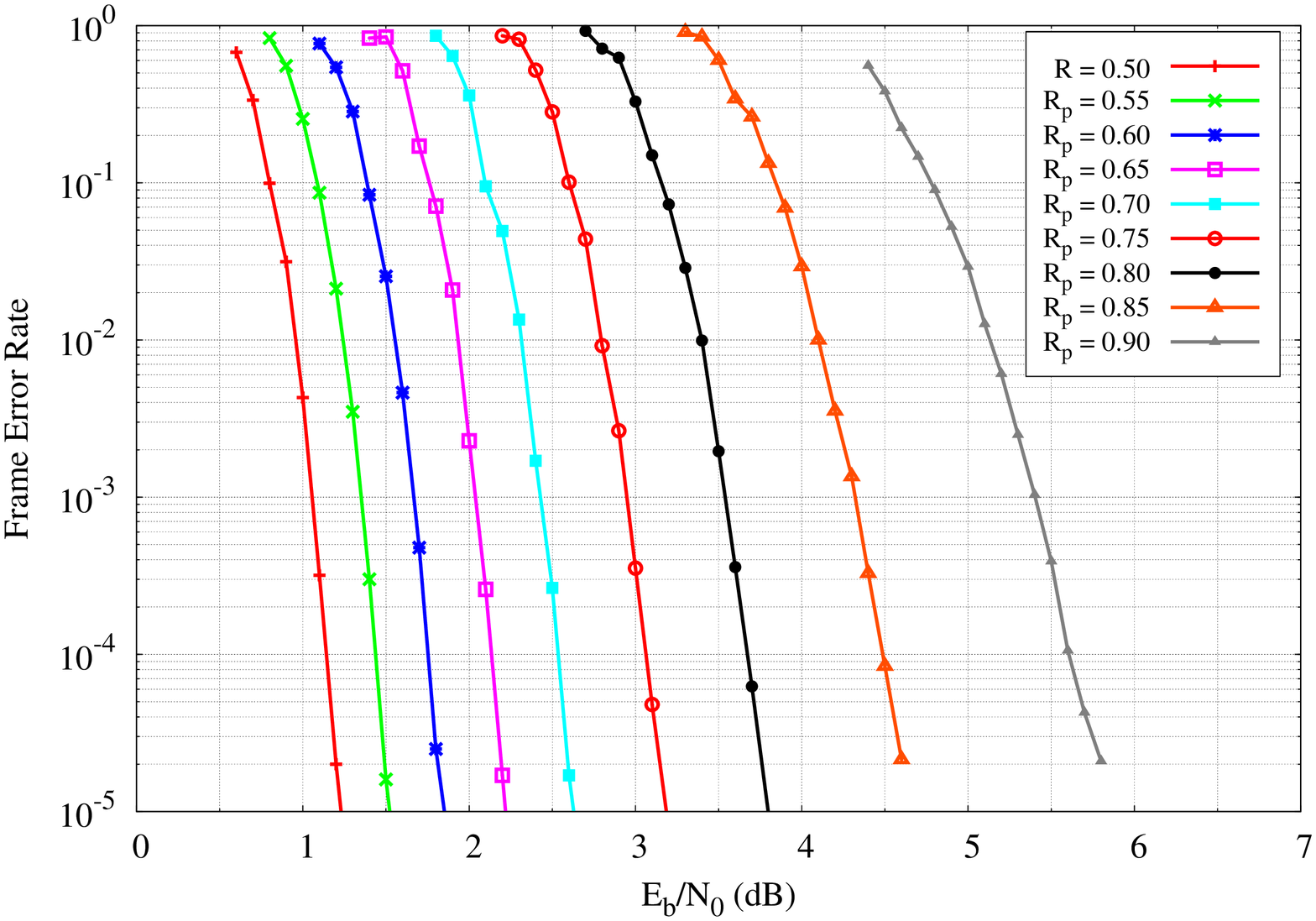}
\vspace{-8mm}\caption{Frame error rate, optimized distributions, $K = 4000$  bits}\label{fig:punct_nbldpc_fer}
\vspace{-3mm}
\end{figure}

Finally, Figure \ref{fig:punct_nbldpc_fer} presents the Frame Error Rate performance of optimized distributions for
finite code lengths. All the codes have binary dimension (number of source bits) equal to 4000 bits (1000
$\gf_{16}$-symbols). The mother code with rate $1/2$ has been constructed by using the Progressive Edge Growth (PEG)
algorithm \cite{PEG}, and punctured patterns have been randomly chosen according to the optimized distribution. It is
very likely that the performance can be further improved by using an optimized design of the puncturing patterns (in
terms of symbol-recoverability steps, cf. discussion in the introduction), especially for high punctured rates.

\section{Conclusions} \label{sec:conclusions}
Except for simple channel models, as for instance the binary erasure channel, it is generally an arduous task to derive
exact density evolution for non-binary LDPC codes. Yet, even if the exact density evolution were known, its inherent
complexity would certainly prohibit its use for optimization purposes.

In this paper we  proposed a density evolution approximation method, by using Monte-Carlo simulation of an infinite
code. The proposed method provides accurate and precise estimates of non-binary ensemble thresholds, and makes possible
the optimization of non-binary codes for a wide range of applications and channel models.

As an application, rate-adaptability solutions for non-binary LDPC codes have been investigated. Puncturing
distributions for regular and irregular codes have been analyzed by using simulated density evolution thresholds of
non-binary LDPC codes over the AWGN channel. For regular codes, we showed that the design of puncturing patterns must
respect different rules depending on the symbol-node degree: punctured bits must be spread over degree-$2$
symbol-nodes, while they must be clustered on symbol-nodes of higher degrees. If the number of punctured bits is
relatively small, spreading punctured bits over degree-$2$ symbol-nodes yields also good performance for irregular
codes. However, such a puncturing distribution could be catastrophic for higher punctured rates, in which case
optimized puncturing distributions are indispensable. Finally, we presented optimized puncturing distributions for
non-binary LDPC codes with small maximum degree, which exhibit a gap to capacity between $0.2$ and $0.5$ dB, for
punctured rates varying from $0.5$ to $0.9$.

 \bibliographystyle{../bib/IEEEbib}
 \bibliography{../bib/MyBiblio,../bib/Zotero}

\end{document}